\pdfoutput=1
\documentclass[aps, twocolumn, floatfix, superscriptaddress, prb]{revtex4-1}
\usepackage{graphicx}
\DeclareGraphicsExtensions{.pdf}
\usepackage{amsmath,amssymb,bbold,bm,color}
\usepackage{float}
\usepackage{epstopdf}
\usepackage{tabularx}
\usepackage{braket}
\usepackage{booktabs}
\usepackage{array}
\usepackage{blindtext}

\usepackage[colorlinks=true,breaklinks=true,linkcolor=blue,anchorcolor=red,citecolor=blue,urlcolor=blue]{hyperref}

\begin{document}

\title{Pseudo-magnetic fields in square lattices}

\author{Junsong Sun}
\affiliation{School of Physics, Beihang University,
Beijing, 100191, China}

\author{Xingchuan Zhu }
\affiliation{Interdisciplinary Center for Fundamental and Frontier Sciences, Nanjing University of Science and Technology, Jiangyin, Jiangsu 214443, P. R. China}

\author{Tianyu Liu}
\email{liuty@sustech.edu.cn}
\affiliation{Shenzhen Institute for Quantum Science and Engineering and Department of Physics, Southern University of Science and Technology (SUSTech), Shenzhen 518055, China}
\affiliation{International Quantum Academy, Shenzhen 518048, China}

\author{Shiping Feng}
\affiliation{ Department of Physics,  Beijing Normal University, Beijing, 100875, China}

\author{Huaiming Guo}
\email{hmguo@buaa.edu.cn}
\affiliation{School of Physics, Beihang University,
Beijing, 100191, China}

\begin{abstract}
We have investigated the effects of strain on two-dimensional square lattices and examined the methods for inducing pseudo-magnetic fields. In both the columnar and staggered $\pi$-flux square lattices, we have found that strain only modulates Fermi velocities rather than inducing pseudo-magnetic fields. However, spatially non-uniform on-site potentials (anisotropic hoppings) can create pseudo-magnetic fields in columnar (staggered) $\pi$-flux square lattices. On the other hand, we demonstrate that strain does induce pseudo-magnetic fields in staggered zero-flux square lattices. By breaking a quarter of the bonds, we clarify that a staggered zero-flux square lattice is topologically equivalent to a honeycomb lattice and displays pseudo-vector potentials and pseudo-Landau levels at the Dirac points.
\end{abstract}

%\pacs{
%  71.10.Fd, % Lattice fermion models (Hubbard model, etc.)
%  03.65.Vf, % Topological phases (quantum mechanics)
%  71.10.-w, % Theories and models of many-electron systems; see also
%            % 67.10.Db Fermion degeneracy in quantum fluids)
%}

\maketitle
%\begin{widetext}
%% \textit{Introduction.-}
\section{Introduction}
Strain engineering has emerged as a powerful tool in condensed matter physics for manipulating the electronic properties of Dirac materials. In graphene, specifically, the application of strain~\cite{castroneto2009} can generate a pseudo-magnetic field~\cite{dejuan2013, vozmediano2010, guinea2010a, guinea2010b, neekamal2013, guinea2008, zhang2014, lantagne2020} that couples to the two-dimensional Dirac electrons in a manner similar to an externally applied magnetic field. Numerous experimental~\cite{levy2010, hsu2020, meng2013, lisy2015, lisy2020, nigge2019, jia2019} and theoretical~\cite{dejuan2012, chang2012, settnes2016, shi2021, liu2022} studies have identified in graphene such strain-induced pseudo-magnetic fields, which give rise to novel transport phenomena such as chiral anomalies~\cite{lantagne2020, shi2021} and quantum oscillations~\cite{liu2022} in the absence of magnetic fields.

Beyond graphene, pseudo-magnetic fields can also be induced in various other materials, including superconducting~\cite{liu2017b, matsushita2018, massarelli2017, nica2018}, magnonic~\cite{PhysRevLett.123.207204,liu2019, liu2021, liu2023, sun2021a, sun2021b}, photonic~\cite{rechtsman2013}, and acoustic~\cite{wen2019, brendel2017, peri2019} materials, as long as they possess a Dirac cone band structure. In the case of two-dimensional materials, strain-induced pseudo-magnetic fields have previously been predicted only in honeycomb-like lattices (e.g., honeycomb~\cite{poli2014, bao2023,PhysRevLett.117.266801,kohler2023nodal}, kagome~\cite{liu2020}, and $\alpha-T_3$\cite{sun2022, filusch2022} lattices), because their lattice geometry inherently guarantees the presence of Dirac cones. However, it remains unknown whether strain can induce pseudo-magnetic fields in non-honeycomb-like lattices.

Square lattices are well-known for exhibiting Dirac cones in the low-energy band structure when each elementary plaquette hosts half a magnetic flux quantum. Due to this unique band structure, the $\pi$-flux square lattices have attracted significant attention and have been extensively studied in both non-interacting~\cite{harris1989, delplace2010, hou2009} and strongly correlated~\cite{PhysRevB.88.075101,zhou2018, davis2019, zhang2020, shaffer2022} regimes. Furthermore, the presence of Dirac cones in the $\pi$-flux square lattices is a prerequisite for the emergence of strain-induced pseudo-magnetic fields. Therefore, it is intriguing and worthwhile to investigate whether a strained $\pi$-flux square lattice can indeed give rise to a pseudo-magnetic field, which may have potential experimental realizations in optical lattices~\cite{aidelsburger2013, miyake2013} or electrical circuits~\cite{lee2018}.

In this manuscript, we investigate the effects of strain on two-dimensional square lattices with and without $\pi$-flux. We examine two different configurations of $\pi$-flux and observe that strain alone does not result in the induction of a pseudo-magnetic field in either case. However, we discover that spatially non-uniform on-site potentials or anisotropic hoppings can serve as alternative sources for generating pseudo-magnetic fields. For the case without $\pi$-flux, we observe that Dirac cones can be produced by introducing staggered hoppings along the $y$ direction. Additionally, we find that strain patterns commonly used in graphene have the ability to induce pseudo-magnetic fields in this system. In the limit case where the weak bonds in the $y$ direction are eliminated, the resulting brick-wall square lattice is topologically equivalent to a honeycomb lattice. Interestingly, the pseudo-magnetic field induced by strain persists in this transformed square geometry. These findings expand the effect of strain to square geometries, further deepening our comprehension of the strain-induced pseudo-magnetic field.

This paper is organized as follows. In Sec.~\ref{sec2}, we introduce the columnar $\pi$-flux square lattice and demonstrate its low-energy Dirac cone band structure. In Sec.~\ref{sec3}, we show that strain only modulates the Fermi velocity of the columnar $\pi$-flux square lattice without inducing a pseudo-magnetic field. In Sec.~\ref{sec4}, we propose that a pseudo-magnetic field can arise when a spatially non-uniform on-site potential is introduced to the columnar $\pi$-flux square lattice. In Sec.~\ref{sec5}, we find that strain cannot induce a pseudo-magnetic field in the staggered $\pi$-flux square lattice, but spatially non-uniform anisotropic hoppings may generate one. In Sec.~\ref{sec6}, we demonstrate that strain can induce a pseudo-magnetic field in the staggered zero-flux square lattice. In Sec.~\ref{sec7}, we reveal that breaking a quarter of bonds in the staggered zero-flux square lattice results in topological equivalence to a honeycomb lattice, exhibiting a strain-induced pseudo-magnetic field. Finally, in Sec.~\ref{sec8}, we provide a summary of our key findings and conclude the paper.

\section{Columnar $\pi$-flux square lattice}
\label{sec2}
\begin{figure}[t]
\centering
\includegraphics[width=8.6cm]{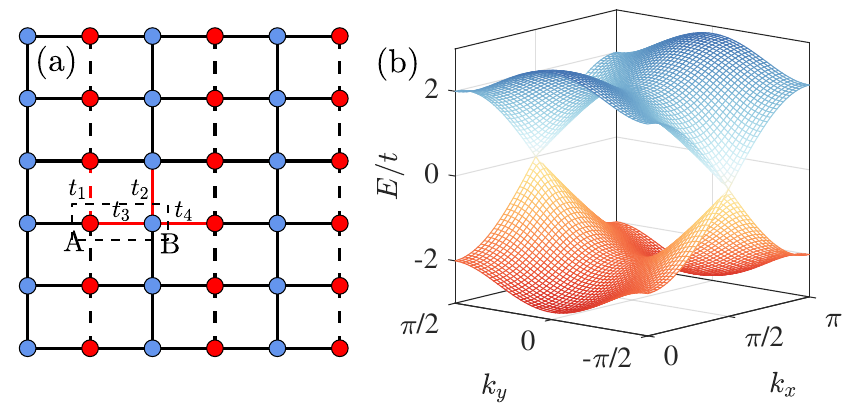}
\caption{(a) Schematic plot of a columnar $\pi$-flux square lattice. Each unit cell consists of two sites, labeled as A (red) and B (blue), respectively. The hopping parameters of the four bonds associated with each unit cell are labeled as $t_{1,2,3,4}$. (b) Band structure of the nearest-neighbor tight-binding model [Eq.~(\ref{H0_col})] defined on the columnar $\pi$-flux square lattice. Here we have adopted $-t_1=t_2=t_3=t_4=t$.
}\label{fig1}
\end{figure}

With half of a magnetic flux quantum threading through each plaquette of a square lattice, the hopping parameters associated with the four edges of the plaquette acquire Aharonov-Bohm phases that sum up to $\pi$. By assigning this $\pi$ phase to $t_1$, the square lattice manifests a columnar pattern with a bipartite unit cell [Fig.~\ref{fig1}(a)]. The corresponding spinless nearest-neighbor tight-binding Hamiltonian is given by:
%
%\begin{align}
%\nonumber
%  H_0&=\sum_{i\in A}\left(-t_1c_i^\dagger c_{i+{\bm \delta}_2}+t_3c_i^\dagger c_{i+{\bm \delta}_1}\right)\\
%  &+\sum_{i\in B}\left(t_2c_i^\dagger c_{i+{\bm \delta}_2}+t_4c_i^\dagger c_{i+{\bm \delta}_1}\right)+{\rm H.c.},
%\end{align}
%
%where $c_i^\dagger$ and $c_i$ denote the creation and annihilation operators, respectively. Furthermore, ${\bm \delta}_1=(1,0)$ and ${\bm \delta}_2=(0,1)$ correspond to the nearest neighbor vectors in the $x$ and $y$ directions. For a more general analysis, we will assign different values for the hopping amplitudes $t_n(n=1,2,3,4)$ on the bonds of each unit cell. In the momentum space and under the basis $\psi_{\bm k}=(c_{\bm k,A},c_{\bm k,B})^{T}$, the Hamiltonian can be written as $H_0=\sum_{\bf k} \psi_{\bm k}^{\dagger}\mathcal{H}_{\bm k}\psi_{\bm k}$ with $\mathcal{H}_{\bm k}$ given by
%
\begin{equation} \label{H0_col}
\begin{split}
{
H_0=\sum_{\bm r} (t_1 a_{\bm r}^\dagger a_{\bm r+\bm\delta_2}+t_2 b_{\bm r}^\dagger b_{\bm r+\bm \delta_2} + t_3 a_{\bm r}^\dagger b_{\bm r}
}
\\
{
+ t_4 a_{\bm r}^\dagger b_{\bm r-2\bm \delta_1})+\text{H.c.},
}
\end{split}
\end{equation}
where $a_{\bm r}$ and $b_{\bm r}$ are the annihilation operators associated with the two sublattices, $\bm \delta_1=(1,0)$ and $\bm \delta_2=(0,1)$ are the nearest-neighbor vectors with the lattice constant set to unity. The hopping parameters [Fig.~\ref{fig1}(a)] associated with each unit cell satisfy $-\text{sgn}(t_1)=\text{sgn}(t_2)=\text{sgn}(t_3)=\text{sgn}(t_4)$. In momentum space and the sublattice basis $\psi_{\bm k}=(a_{\bm k},b_{\bm k})^{T}$, the Hamiltonian [Eq.~(\ref{H0_col})] can be written as $H_0=\sum_{\bm k} \psi_{\bm k}^{\dagger}\mathcal{H}_{\bm k}\psi_{\bm k}$ with the kernel given by
\begin{align}\label{H0k_col}
\mathcal{H}_{\bm k} &=\left\lbrack \begin{array}{cc}
2t_1 \cos \left(k_y \right) & t_3 +t_4 e^{2\mathrm{i}k_x } \\
t_3 +t_4 e^{-2\mathrm{i}k_x }  & {2t}_2 \cos \left(k_y \right)
\end{array}\right\rbrack.
\end{align}
Taking $-t_1=t_2=t_3=t_4=t$, the band structure of $\mathcal{H}_{\bm k}$ reads $E_{\bm k}=\pm2t\sqrt{\cos^2(k_x)+\cos^2(k_y)}$, which exhibits two Dirac points at ${\bm K}^\xi=\left(\frac{\pi}2, \xi\frac{\pi}2\right)$ with $\xi=\pm$ [see Fig.\ref{fig1}(b)]. We expand $\mathcal{H}_{\bm k}$ in the vicinity of ${\bm K}^\xi$ and obtain the corresponding low-energy effective Hamiltonian
%%
%\begin{equation}
%  H_{\bm q}^\xi =\left\lbrack \begin{array}{cc}
%-2\xi t_1q_y  & t_3-t_4-2t_4{\rm i}q_x\\
%t_3-t_4+2t_4{\rm i}q_x & -2\xi t_2q_y
%\end{array}\right\rbrack.
%\end{equation}
%%
%Here, ${\bm q}={\bm k}-{\bm K}^\xi$ represents the small momentum deviation from the Dirac point. Specifically, when $t_1=-t$ and $t_2=t_3=t_4=t$, the effective Hamiltonian $H_{\bm q}^\xi$ takes the form
%
\begin{align}\label{H0q_col}
  H_{\bm q}^\xi=2t\left\lbrack \begin{array}{cc}
\xi q_y  & -{\textrm{i}}q_x \\
{\textrm{i}}q_x  & -\xi q_y
\end{array}\right\rbrack =2t\left(\xi\sigma_z q_y+\sigma_y q_x\right),
\end{align}
where $\sigma_y$ and $\sigma_z$ are Pauli matrices.
As a consequence, a linear dispersion relation with isotropic properties emerges in the vicinity of the Dirac points, expressed as $\varepsilon_{\bm q}=\pm2t\sqrt{q_x^2+q_y^2}$.

\section{Strain-modulated Fermi velocity in the columnar $\pi$-flux square lattice}
\label{sec3}
We now study the strain effects in the columnar $\pi$-flux square lattice. When the strain is applied, it deforms the lattice and modulates the hopping parameters to
\begin{equation}
  t_n=t+\delta t_n,
\end{equation}
where $\delta t_n$ denotes the correction to the $n$th hopping parameter. Any strain, regardless of its space dependence, can be characterized by a displacement field ${\bm U}(\bm R)$. The variation $\delta t_n$ can be approximated in terms of the strain tensor $u_{ij}= (\partial_iU_j+\partial_jU_i)/2$ and the corresponding bond vector ${\bm \delta}_n$ as
\begin{equation}\label{dtn}
%\delta t_n=-\frac{\beta t}{a_0^2}{\bm\delta}_n\cdot{\bm u}\cdot{\bm \delta_n},
{
\delta t_n=-\beta t{\bm\delta}_n\cdot{\bm u}\cdot{\bm \delta_n},
}
\end{equation}
%
%$\beta=-\partial\ln t({\bm \delta}_n)/\partial \ln|{\bm \delta}_n|_{|{\bm\delta}_n|\textcolor{red}{=1}}$
where $\beta$ is referred to as the Gr\"{u}neisen parameter~\cite{vozmediano2010}. For the columnar $\pi$-flux square lattice, the strain-modulated hopping parameters thus read
%%
%\begin{align}
%\nonumber
%  &t_{1/2}=\pm t(1-\frac{\beta}{a^2}\bm{\delta}_2\cdot\bm{u}\cdot\bm{\delta}_2)\\
%  \nonumber
%  &t_{3/4}=~~t(1-\frac{\beta}{a^2}\bm{\delta}_1\cdot\bm{u}\cdot\bm{\delta}_1).
%\end{align}
%
%As both $t_1$ and $t_2$ (as well as $t_3$ and $t_4$) are along the direction of ${\bm \delta}_1$(${\bm \delta}_2$), they are affected similarly by any deformation described by the displacement function ${\bm U}({\bm R})$. Consequently, we have $t_1= -t_2$ and $t_3= t_4$ regardless of the form of ${\bm U}({\bm R})$. With the correction to the hopping amplitude given by Eq.(7), the effective Hamiltonian in Eq.(3) becomes:
%
\begin{equation}\label{t_mod}
\begin{split}
{
t_{1,2}=\pm t(1-\beta\bm{\delta}_2\cdot\bm{u}\cdot\bm{\delta}_2),
}
\\
{
t_{3,4}=t(1-\beta\bm{\delta}_1\cdot\bm{u}\cdot\bm{\delta}_1),
}
\end{split}
\end{equation}
where the plus (minus) sign in the first equation corresponds to $t_2$ ($t_1$). According to Eq.~(\ref{t_mod}), we always have $t_1= -t_2$ and $t_3= t_4$ regardless of the form of ${\bm U}({\bm R})$. Plugging Eq.~(\ref{t_mod}) into Eq.~(\ref{H0k_col}) and expanding around $\bm K^\xi$, we find that the low-energy effective Hamiltonian [Eq.~(\ref{H0q_col})] is adapted to
\begin{equation}
{
H_{\bm q}^\xi=2t[\xi\sigma_z (1-\beta\bm{\delta}_2\cdot\bm{u}\cdot\bm{\delta}_2)q_y +\sigma_y (1-\beta\bm{\delta}_1\cdot\bm{u}\cdot\bm{\delta}_1)q_x],
}
\end{equation}
{which only incorporates modulation of the Fermi velocity instead of induction of a pseudo-magnetic field. For spatially uniform strain, the Fermi velocity remains constant and exhibits anisotropy when $\bm{\delta}_1\cdot\bm{u}\cdot\bm{\delta}_1 \neq \bm{\delta}_2\cdot\bm{u}\cdot\bm{\delta}_2$. For non-uniform strain, the Fermi velocity in general exhibits anisotropy and spatial inhomogeneity simultaneously.}

%From the above equation, we can conclude that applying strain to the $\pi$-flux square lattice only leads to the appearance of coefficients in front of $q_x$ and $q_y$. These coefficients modify the Fermi velocities instead of creating a pseudo-magnetic field. Since the strain tensor for uniform strain remains constant, the coefficients are also constant. As a result, uniform strain causes an anisotropic Dirac cone where the Fermi velocity differs along the $q_x$ and $q_y$ directions. Conversely, non-uniform strain involves strain tensor components that vary with coordinates, leading to inhomogeneous Fermi velocities in both the $q_x$ and $q_y$ directions.

\begin{figure}[t]
\centering
\includegraphics[width=8.8cm]{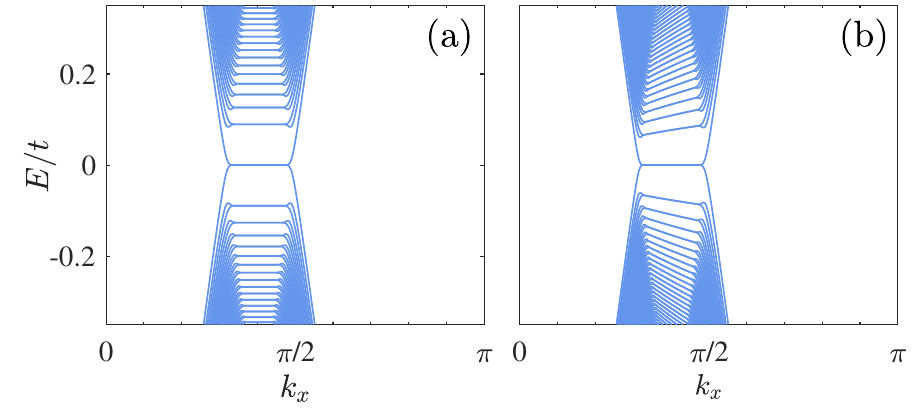}
\caption{Band structure of the nearest-neighbor tight-binding model on the columnar $\pi$-flux square lattice with an ordinary magnetic field. (a) Flat Landau levels. (b) Dispersive Landau levels in the presence of a non-uniform uniaxial strain. The strain results from the displacement field ${\bm U}=(0,\frac{c}{2\beta}y^2)$ with strength $c/c_{\text{max}}=0.5$ ($c_{\text{max}}=1/L_y$). The lattice used in the calculation has a finite width ($L_y=600$) in the $y$ direction and is infinite along the $x$ direction.
}\label{fig2}
\end{figure}

It is well known that strain in graphene can induce both an inhomogeneous Fermi velocity and a pseudo-magnetic field, which together give rise to dispersive pseudo-Landau levels~\cite{liu2022, lantagne2020}. However, in the case of the columnar $\pi$-flux square lattice, non-uniform strain only generates an inhomogeneous Fermi velocity. It is thus anticipated that an external magnetic field produces dispersive Landau levels. {In this scenario, the magnetic field $\bm B=B\hat z$ is responsible for the Landau quantization $E_n=\sqrt{2neB\hbar v_xv_y}$ as illustrated in Fig.~\ref{fig2}(a), while the non-uniform Fermi velocity $\bm v=(v_x,v_y)$ becomes $\bm k$ dependent when Fourier-transformed into the momentum space, resulting in the dispersion. Our claim is numerically substantiated with a displacement field ${\bm U}=(0,\frac{c}{2\beta}y^2)$ through exact diagonalization. Indeed, we find that the initially flat Landau levels [Fig.~\ref{fig2}(a)] become dispersive [Fig.~\ref{fig2}(b)] due to the non-uniform Fermi velocity modulated by the strain.}
%Next we consider a non-uniform uniaxial strain with the displacement function ${\bm U}=(0,\frac{c}{2\beta}y^2)$. By considering both the aforementioned strain and a real magnetic field, and diagonalizing the resulting tight-binding model, we can directly obtain the energy spectrum. As depicted in Fig.\ref{fig2}(b), the initially flat Landau levels become dispersive due to the non-uniform Fermi velocity induced by the strain.

%

Previously, there has been a debate regarding the origin of the dispersive pseudo-Landau levels in graphene. Recent findings~\cite{liu2022} have revealed that the dispersion arises from both the strain-modulated inhomogeneous Fermi velocity~\cite{lantagne2020, oliva2020} and the strain-induced non-uniform pseudo-magnetic field~\cite{shi2021}. In fact, these two effects cannot be separated for strong strain~\cite{liu2022}. In contrast, in the columnar $\pi$-flux square lattice, strain independently affects the Fermi velocity and thus can serve as a more agile tuning knob of the electronic structure.
%Recent findings have definitively established that the dispersion of the pLLs is indeed a bulk effect. Here, in the columnar $\pi$-flux square lattice, the presence of strain allows for the independent creation of inhomogeneous Fermi velocities. This presents a unique opportunity to validate its key role in generating the observed dispersion in pLLs, thereby enhancing our comprehension of pLLs at a deeper level.

\section{On-site potential induced pseudo-magnetic field in the columnar $\pi$-flux square lattice}
\label{sec4}
\begin{figure}[t]
\centering
\includegraphics[width=8.8cm]{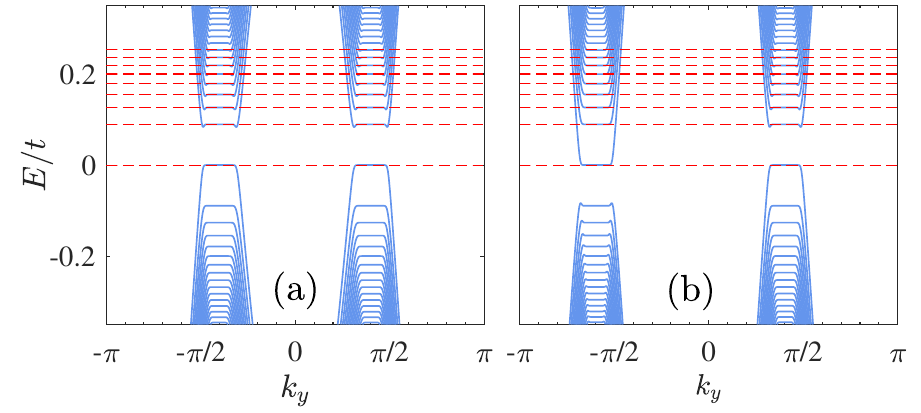}
\caption{The low-energy band structure of the nearest-neighbor tight-binding model of the columnar $\pi$-flux square lattice. (a) Pseudo-Landau levels induced by the engineered non-uniform on-site potential [Eq.~(\ref{pot})]. (b) Landau levels arising from a real magnetic field $B=U_0$. The red curves in (a) and (b) represent the analytical Landau levels [Eq.~(\ref{LL_col})].
}\label{fig3}
\end{figure}

In Sec.~\ref{sec3}, we have shown that the strain-modulated hopping parameters are unable to generate a pseudo-magnetic field on the columnar $\pi$-flux square lattice. Nevertheless, upon analyzing the structure of the low-energy effective Hamiltonian [Eq.~(\ref{H0q_col})], it is observed that a non-uniform on-site potential, which varies with the $x$ coordinate, has the ability to induce a vector potential. Explicitly, the  potential reads
%%
%\begin{equation}
%H_1=2U_0(\sum_{i\in A}x_ic_i^\dagger c_i-\sum_{i\in B}x_ic_i^\dagger c_i),
%\end{equation}
%%
%
\begin{equation}\label{pot}
{
H_1=2tU_0\sum_{\bm r} (\bm r \cdot \hat x) (a_{\bm r}^\dagger a_{\bm r} - b_{\bm r}^\dagger b_{\bm r})},
\end{equation}
where $U_0$ characterizes the strength of the potential. The total Hamiltonian now becomes $H_{tot}=H_0+H_1$, whose low-energy effective Hamiltonian [cf., Eq.~(\ref{H0q_col})] writes as
\begin{equation}\label{Hq_col}
\begin{split}
H_{\bm q}^\xi&=2t \begin{bmatrix}
\xi q_y+U_0x  & -{\textrm{i}}q_x
\\
{\textrm{i}}q_x  & -\xi q_y-U_0x
\end{bmatrix}
\\
&=2t\left[\xi\sigma_z (q_y+\xi U_0x)+\sigma_y q_x\right].
\end{split}
\end{equation}
It is worth noting that $H_1$ is an artificial term, similar to an electric field, but experienced oppositely by the two sublattices. The Hamiltonian [Eq.~(\ref{Hq_col})] exhibits a vector potential $\bm A^{\xi}=(0, \xi U_0x)$, which has opposite signs at the two Dirac points. Solving the eigenvalue problem of Eq.~(\ref{Hq_col}) yields the pseudo-Landau levels (see Appendix~\ref{a1} for details)
\begin{equation}\label{LL_col}
E_n=\pm2t\sqrt{2U_0n},\qquad n=0,1,2,\cdots,
\end{equation}
which exhibit a $\sqrt n$ dependence, similar to that in strained graphene. The analytical dispersion [Eq.~(\ref{LL_col})] can be further verified by diagonalizing the corresponding tight-binding model on the columnar $\pi$-flux square lattice, with open (periodic) boundary condition along the $x$ $(y)$ direction. As shown in Fig.~\ref{fig3}(a), the analytical pseudo-Landau levels and the numerical bands match quite well with each other near the Dirac points. For comparison, we also plot in Fig.~\ref{fig3}(b) the Landau levels induced by a real magnetic field with the same strength (i.e., $B=U_0$). While the pseudo-Landau levels at the two Dirac points are linked by the time-reversal symmetry, the two sets of Landau levels in Fig.~\ref{fig3}(b) are related to each other by the inversion symmetry.

\section{Staggered $\pi$-flux square lattice}
\label{sec5}
\begin{figure}[t]
\centering
\includegraphics[width=8.6cm]{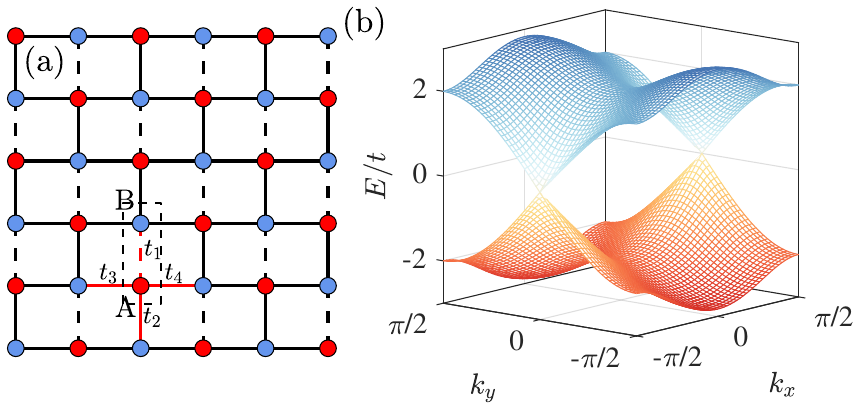}
\caption{(a) Schematic plot of a staggered $\pi$-flux square lattice. Each unit cell consists of two sites, labeled as A (red) and B (blue), respectively. The hopping parameters of the four bonds associated with each unit cell are labeled as $t_{1,2,3,4}$. (b) Band structure of the tight-binding model [Eq.~(\ref{H0_sta})] defined on the staggered $\pi$-flux square lattice. Here we have adopted $-t_1=t_2=t_3=t_4=t$.
}\label{fig4}
\end{figure}

We have so far focused on the columnar $\pi$-flux square lattice in Secs.~\ref{sec2}-\ref{sec4}. In the present section, we analyze a different $\pi$-flux square lattice which illustrates a staggered pattern [Fig.~\ref{fig4}(a)]. We will examine the strain effects and study the possible induction of pseudo-magnetic fields.

The tight-binding model of a staggered $\pi$-flux square lattice reads
%%
%\begin{align}
%\nonumber
%  H_0&=\sum_{i\in A}\left(-t_1c_i^\dagger c_{i+{\bm \delta}_2}+t_2c_i^\dagger c_{i-{\bm \delta}_2}\right)\\
%  &+\sum_{i\in A}\left(t_3c_i^\dagger c_{i-{\bm \delta}_1}+t_4c_i^\dagger c_{i+{\bm \delta}_1}\right)+{\rm H.c.}.
%\end{align}
%%
%
\begin{equation} \label{H0_sta}
\begin{split}
{
H_0=\sum_{\bm r} (t_1 a_{\bm r}^\dagger b_{\bm r} + t_2 a_{\bm r}^\dagger b_{\bm r-2\bm \delta_2} + t_3 a_{\bm r}^\dagger b_{\bm r-\bm \delta_1 -\bm \delta_2}
}
\\
{
+ t_4 a_{\bm r}^\dagger b_{\bm r+\bm \delta_1 -\bm \delta_2}) +\text{H.c.}.
}
\end{split}
\end{equation}
Performing Fourier transform, we find in the sublattice space the following Bloch Hamiltonian
\begin{equation}\label{H0k_sta}
  {\cal H}_{\bm k} =\left\lbrack \begin{array}{cc}
0 & f_{\bm k} \\
  f_{\bm k}^* & 0
\end{array}\right\rbrack,
\end{equation}
where $f_{\bm k}=t_1+t_2e^{-{\rm i}2k_y } +t_3 e^{-{\rm i}\left(k_x +k_y \right)} +t_4 e^{{\rm i}\left(k_x-k_y \right)}$. Taking $-t_1=t_2=t_3=t_4=t$, we find that $\mathcal H_{\bm k}$ exhibits two Dirac points at {$\bm K^\xi=(\xi\tfrac{\pi}{2},0)$} [see Fig.\ref{fig4}(b)], in the vicinity of which, the low-energy effective Hamiltonian is obtained through linearization as
%The uniform system has two Dirac points are located at $K^{+}=\left(\frac{\pi}2, 0\right)$ and $K^{-}=\left(\frac{\pi}2, \pi\right)$, respectively. Expanding ${\cal H}_{\bm k}$ around $K^{\pm}$, we can obtain the low-energy effective Hamiltonian as follows:
%\begin{align}
%\nonumber
%  H_{\bm q}^\xi =&\sigma_x\left[t_1 +t_2 -\xi q_x \left(t_3 +t_4 \right)-\xi q_y \left(t_3 -t_4 \right)\right]\\
%  +&\sigma_y\left[2t_2 q_y +\xi(t_3 -t_4) \right].
%\end{align}
%When $t_1=-t, t_2=t_3=t_4=t$, we have $H_{\bm q}^\xi =-2t\left\lbrack \sigma_x \xi q_x -\sigma_y q_y \right\rbrack$.
%The dispersion of the Dirac cone is $\varepsilon_{\bm q}=\pm2t\sqrt{q_x^2+q_y^2}$, which is exactly the same with that under the Landau gauge.
%
\begin{equation}
H_{\bm q}^\xi =-2t (\sigma_x \xi q_x -\sigma_y q_y),
\end{equation}
whose spectrum $\varepsilon_{\bm q}=\pm2t\sqrt{q_x^2+q_y^2}$ is exactly the same as that of the columnar $\pi$-flux square lattice.

%

%Next, we consider the effect of strain with the strain tensor $\bm{u}$. Similarly, the modified hopping amplitudes are given by
%\begin{align}
%\nonumber
%  &t_{1/2}=\pm t(1-\frac{\beta}{a^2}\bm{\delta}_2\cdot\bm{u}\cdot\bm{\delta}_2)\\
%  &t_{3/4}=~~t(1-\frac{\beta}{a^2}\bm{\delta}_1\cdot\bm{u}\cdot\bm{\delta}_1).
%\end{align}
%Substituting the above equation into Eq(14) yields the low-energy effective Hamiltonian
{We next consider the effect of strain in the staggered $\pi$-flux square lattice. The strain can be incorporated using the same hopping modulation  [Eq.~(\ref{t_mod})]. Plugging Eq.~(\ref{t_mod}) into Eq.~(\ref{H0_sta}), performing Fourier transform, and linearizing in the vicinity of the Dirac points, we find the low-energy effective Hamiltonian}
\begin{equation}
\begin{split}
{
H_{\bm q}^\xi=-2t\xi\sigma_x(1-\beta\bm{\delta}_1\cdot\bm{u}\cdot\bm{\delta}_1)q_x
}
%\\
{
+2t\sigma_y(1-\beta\bm{\delta}_2\cdot\bm{u}\cdot\bm{\delta}_2)q_y.
}
\end{split}
\end{equation}
It is apparent that applying non-uniform strain only results in an inhomogeneous Fermi velocity and cannot produce a pseudo-magnetic field. This observation is consistent with the strain effect in the columnar $\pi$-flux square lattice.

\begin{figure}[t]
\centering
\includegraphics[width=8.8cm]{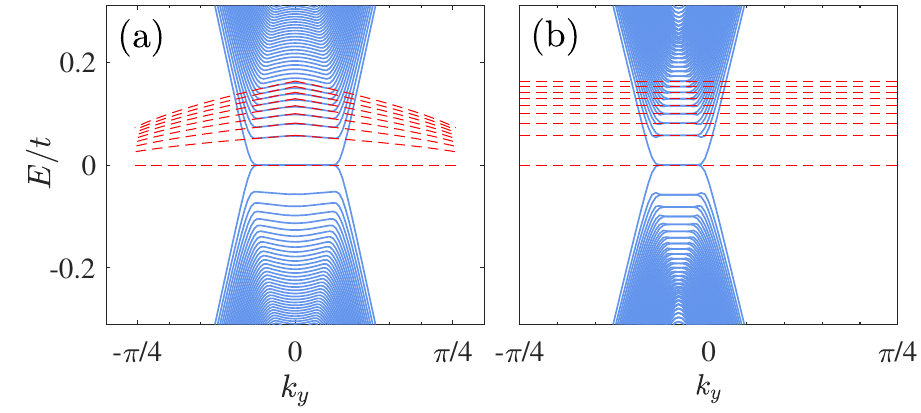}
\caption{The low-energy band structure of the nearest-neighbor tight-binding model on the staggered $\pi$-flux square lattice. (a) Pseudo-Landau levels induced by the engineered hopping parameters [Eq.~(\ref{t_eng})]. The anisotropic hoppings are characterized by $c=0.5c_{\text{max}}$, where $c_{\text{max}}=1/L_x$. (b) Landau levels arising from a real magnetic field $B=c/2$. In both panels, the red curves represent the analytical Landau levels [Eq.~(\ref{LL_sta})] and the system size is $L_x=600$ (infinite) in the $x$ ($y$) direction.
}\label{fig5}
\end{figure}

While a pseudo-magnetic field cannot be produced by strain, it can be artificially created through engineering the hopping parameters. One such example reads
\begin{equation}\label{t_eng}
t_1=-t,\ t_2=t_4=t,\ t_3=t(1-cx),
\end{equation}
where $c$ characterizes the inhomogeneous anisotropy of hoppings. The resulting low-energy effective Hamiltonian becomes
\begin{equation}\label{Hq_sta}
  H_{\bm q}^\xi=2t\left[-\left(1-\frac{c}{2}x\right)\xi q_x\sigma_x+\left(q_y -\xi\frac{c}{2}x\right)\sigma_y\right].
\end{equation}
Clearly, a pseudo-vector potential $\bm A^\xi=(0, \xi\tfrac{c}{2}x)$ with opposite signs $\xi=\pm$ is created at the two Dirac points, resulting in a uniform pseudo-magnetic field in the $z$ direction. Solving the eigenvalue problem of Eq.~(\ref{Hq_sta}) yields the pseudo-Landau levels (see Appendix~\ref{a2} for details)
\begin{equation}\label{LL_sta}
E_n^\xi=\pm2t\sqrt{n{|c|}(1-\xi q_y)},\qquad n=0,1,2,\cdots.
\end{equation}
{where the dispersion of the pseudo-Landau levels is originated from the combined effect of the non-uniform pseudo-magnetic field and Fermi velocity. The analytically derived pseudo-Landau levels [Eq.~(\ref{LL_sta})] well match the numerical energy bands obtained through exact diagonalization of the nearest-neighbor tight-binding model on the staggered $\pi$-flux square lattice [Fig.~\ref{fig5}(a)]. The dispersion of the pseudo-Landau levels makes them stand out from the regular flat Landau levels [Fig.~\ref{fig5}(b)] produced by a real magnetic field.}

%It can be observed from the above equation that the pLLs are dispersive, which is a result of the combined effect of the pseudo-magnetic field and the non-uniform Fermi velocity. The pLLs obtained from both the analytical formula and the tight-binding calculations are plotted in Fig.\ref{fig5}(a), which are in good agreement with each other near the Dirac points. Compared to the dispersive pLLs, the Landau levels induced by a real magnetic field are strictly flat, as shown in Fig.\ref{fig5}(b).

%

\begin{figure}[t]
\centering
\includegraphics[width=8.8cm]{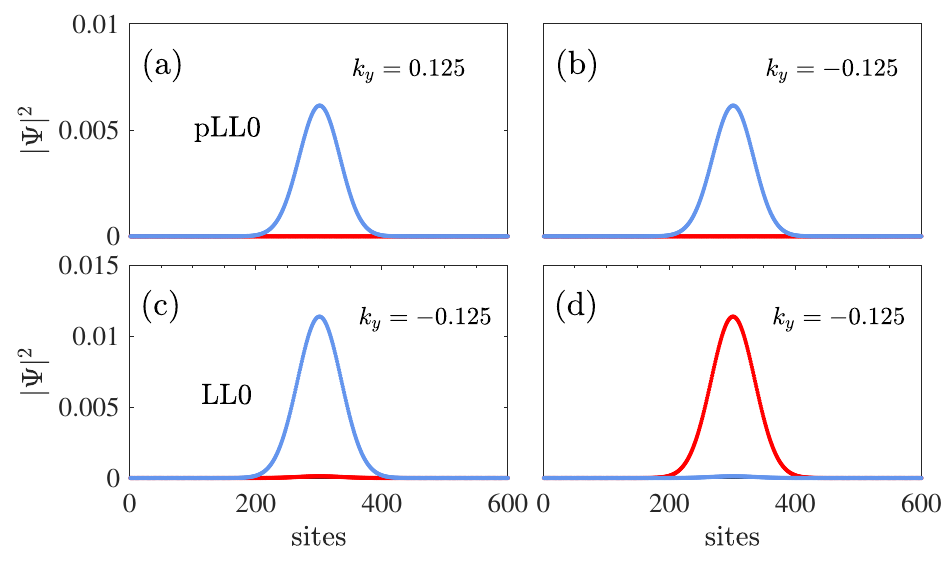}
\caption{The distributions of the wave functions of zeroth (pseudo-) Landau levels at different momenta. (a) Zeroth pseudo-Landau level at {$k_y=0.125$}. (b) Zeroth pseudo-Landau level at {$k_y=-0.125$}. (c, d) Two degenerate sectors of the zeroth Landau levels at {$k_y=-0.125$}. The degeneracy emerges because the two Dirac cones, whose Dirac points are located at $\bm K^\xi=(\xi\tfrac{\pi}{2},0)$, overlap in the vicinity of $k_y=0$ when projected along the $x$ direction. For all panels, the red (blue) curves represent the distribution on the A (B) sublattice. Here we set $c=0.5c_{\text{max}}$.
}\label{fig6}
\end{figure}

{We mention that the zeroth pseudo-Landau levels [Eq.~(\ref{LL_sta})] generated by engineering the hopping parameters [Eq.~(\ref{t_eng})] also exhibit sublattice polarization, and are different from the ordinary zeroth Landau levels that distribute on both sublattices. To substantiate this claim, we plot the distributions of the wave functions at the two valleys (i.e., Dirac cones). The zeroth pseudo-Landau levels are only distributed on the B sublattice for $c>0$ at both valleys [Figs.~\ref{fig6}(a) and~\ref{fig6}(b)]. In contrast, the zeroth Landau levels can appear on either A or B sublattices [Figs.~\ref{fig6}(c) and~\ref{fig6}(d)], depending on which valley is examined. It is worth noting that the sublattice polarization flips when the sign of $c$ is reversed.}

\section{strain-induced pseudo-magnetic field in the staggered zero-flux square lattice}
\label{sec6}
{In Sec.~\ref{sec5}, we have shown that strain only modulates the Fermi velocity of the staggered $\pi$-flux square lattice. In the present section, we demonstrate that strain may induce a pseudo-magnetic field if the flux is removed [i.e., $\text{sgn}(t_1)=\text{sgn}(t_2)=\text{sgn}(t_3)=\text{sgn}(t_4)$] from the staggered $\pi$-flux square lattice. The circumvention of the negative hopping should also render the lattice more experimentally accessible.}

{Removing the flux, the resulting staggered zero-flux square lattice can still be characterized by the tight-binding Hamiltonian [Eq.~(\ref{H0_sta})] and the Bloch Hamiltonian [Eq.~(\ref{H0k_sta})] of the staggered $\pi$-flux square lattice [Fig.~\ref{fig4}(a)], except that we now take $t_1=rt$ and $t_2=t_3=t_4=t$ rather than $-t_1=t_2=t_3=t_4=t$. For the ratio $0\leq r<1$, we find for the Bloch Hamiltonian [Eq.~(\ref{H0k_sta})] two gapless points at ${\bm K}^\xi=[\xi\arccos(-\frac{1+r}{2}),0]$. Expanding the Bloch Hamiltonian [Eq.~(\ref{H0k_sta})] in the vicinity of ${\bm K}^\xi$ yields a low-energy effective Hamiltonian
}
%
%It is worth noting that if the magnitude of $t_1$ is smaller than that of $t_2$, $t_3$, and $t_4$, the resulting band structure will still exhibit Dirac points. Hence, it is possible to avoid using negative hopping amplitude $t_1$ and, consequently, the associated $\pi$ flux as well. This setup should be more readily achievable experimentally. In the following study, we examine the scenario where $t_1=rt (0<r<1)$ and $t_2=t_3=t_4=t$. Here, we utilize the same configuration of $t_{\alpha} (\alpha=1-4)$ as depicted in Fig.\ref{fig4}(a). After scaling down $t_1$ by a factor of $r$, the dispersion transitions from quadratic to linear, and the Dirac points are positioned at ${\bm K}^\xi=[\xi\arccos(-\frac{1+r}{2}),0]$. Consequently, expanding ${\cal H}_{\bm k}$ near ${\bm K}^\xi$ yields an effective Hamiltonian in the vicinity of the Dirac points, which is given by
%\begin{align}
%\nonumber
%  H_{\bm q}^\xi &=\sigma_x \left\lbrace-\xi\frac{\sqrt{(3+r)(1-r)}}{2}\left[(t_3+t_4)q_x+(t_3-t_4)q_y\right]\right.\\ \nonumber
%  &\left.+rt_1+t_2-\frac{(1+r)}{2}(t_3+t_4)\right\rbrace\\ \nonumber
%  &+\sigma_y\left\lbrace-\frac{(1+r)}{2}\left[(t_3-t_4)q_x+(t_3+t_4)q_y\right]+2t_2q_y\right.\\
%  &\left.+\xi\frac{\sqrt{(3+r)(1-r)}}{2}\left[(t_3-t_4)-(t_3+t_4)q_x\right]q_y\right\rbrace.
%\end{align}
%
\begin{equation}\label{H0q_sta0}
{
H_{\bm q}^\xi = -\sigma_xt \xi p q_x +\sigma_y t \left[ (1-r)q_y  -\xi p q_xq_y \right],
}
\end{equation}
where we define $p=\sqrt{(3+r)(1-r)}$ for transparency. We note that Eq.~(\ref{H0q_sta0}) is still a Dirac Hamiltonian when ignoring the $O(q_xq_y)$ term.

\begin{figure}[t]
\centering
\includegraphics[width=8.8cm]{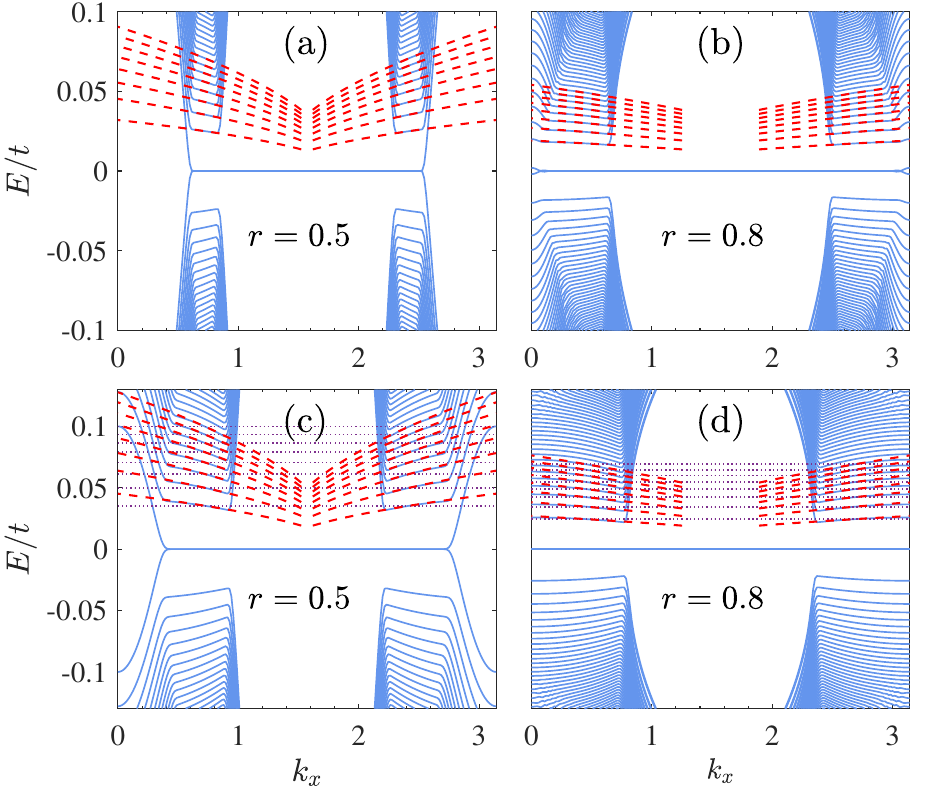}
\caption{The low-energy band structure of the nearest-neighbor tight-binding model on a staggered zero-flux square lattice under different strain patterns and $r$ values. (a) Uniaxial strain, $r=0.5$. (b) Uniaxial strain, $r=0.8$. (c) Triaxial strain, $r=0.5$. (d) Triaxial strain, $r=0.8$. In all panels, the red dashed curves plot the analytical pseudo-Landau levels [Eq.~(\ref{pLL_uni}) for the uniaxial strain and Eq.~(\ref{pLL_tri}) for the triaxial strain]. The width of the strip is $L_y=600$, and the strain strength is $c=0.5c_{\text{max}}$.
}\label{fig7}
\end{figure}

We now check whether such Dirac cones can be Landau-quantized by strain. By incorporating the corrections to the hopping parameters [Eq.~(\ref{t_mod})], we can derive the low-energy effective Hamiltonian in the presence of strain as
\begin{equation}
\begin{split}
{
H_{\bm q}^\xi = \sigma_x t \Big\{-\xi p \left(1-\beta \bm{\delta}_1\cdot\bm{u}\cdot\bm{\delta}_1\right)q_x
}
\\
{
+(1+r) \beta \left(\bm{\delta}_1\cdot\bm{u}\cdot\bm{\delta}_1-\bm{\delta}_2\cdot\bm{u}\cdot\bm{\delta}_2\right) \Big\}
}
\\
{
+\sigma_yt \Big\{ [{(1-r)}+(1+r)\beta\bm{\delta}_1\cdot\bm{u}\cdot\bm{\delta}_1
}
\\
{
-2\beta\bm{\delta}_2\cdot\bm{u}\cdot\bm{\delta}_2]q_y -\xi p \left(1- \beta \bm{\delta}_1\cdot\bm{u}\cdot\bm{\delta}_1\right) q_xq_y \Big\},
}
\end{split}
\end{equation}
which indicates that a uniform pseudo-magnetic field can be induced when $\bm{\delta}_1 \cdot \bm{u} \cdot \bm{\delta}_1 - \bm{\delta}_2 \cdot \bm{u} \cdot \bm{\delta}_2$ is proportional to the $y$ coordinate. This requirement can be fulfilled by a non-uniform uniaxial strain characterized by the displacement field ${\bm U}=(0,\frac{c}{2\beta}y^2)$. Under this strain, the low-energy effective Hamiltonian can be expressed as
\begin{equation}\label{Heff_uni}
\begin{split}
{
H_{\bm q}^\xi = t\left\lbrace\sigma_x\left[-\xi p q_x-(1+r)cy\right]\right.
}
\\
{
\left.+\sigma_y\left[\left(1-r-2cy\right)q_y-\xi p q_xq_y\right]\right\rbrace.
}
\end{split}
\end{equation}
Solving the eigenvalue problem of Eq.~(\ref{Heff_uni}) yields the following strain-induced dispersive pseudo-Landau levels (see Appendix~\ref{a3} for details)
\begin{align}\label{pLL_uni}
{
E_n^\xi=\pm t\sqrt{2(1-r)}\sqrt{n|c|\left(\xi p q_x+1+r\right)},
}
\end{align}
which exhibit a good match to the numerical band structure obtained by diagonalizing the nearest-neighbor tight-binding model that incorporates the strain effect [Figs.~\ref{fig7}(a) and~\ref{fig7}(b)]. From Eq.~(\ref{pLL_uni}), it can be inferred that the interval between the pseudo-Landau levels decreases as $r$ increases from $0$ to $1$. Specifically, when $r=1$, the pseudo-Landau levels collapse, because the ordinary square lattice is restored and the low-energy dispersion becomes quadratic.

A pseudo-magnetic field can also be generated by applying a triaxial strain, which is characterized by the displacement field ${\bm U}(x,y)=\frac{c}{\beta}(2xy,x^2-y^2)$. According to Eq.~(\ref{t_mod}), the hopping parameters are modulated by the triaxial strain as
\begin{equation} \label{t_mod_tri}
\begin{split}
t_1=rt(1+2cy), t_2=t(1+2cy),
%\\
t_3=t_4=t(1-2cy).
\end{split}
\end{equation}
The low-energy effective Hamiltonian can be directly obtained and read
\begin{equation} \label{Heff_tri}
\begin{split}
{
H_{\bm q}^\xi=t\sigma_x\left[-\xi {p}(1-2cy)q_x+(1+r)4cy\right]
}
\\
{
+t\sigma_y\left[1-r-\xi pq_x+2c(3+r)y\right]q_y.
}
\end{split}
\end{equation}
We first consider a simplified solution to the eigenvalue problem of Eq.~(\ref{Heff_tri}) by neglecting the small terms $O(cq_x)$, $O(cq_y)$, and $O(q_xq_y)$. Afterwards, Eq.~(\ref{Heff_tri}) is reduced to a minimally coupled (i.e., Peierls-substituted) Dirac Hamiltonian
\begin{align} \label{Heff_tri_simp}
H_{\bm q}^\xi &= t\left\lbrace\sigma_x\left[-\xi pq_x+(1+r)4cy\right]+\sigma_y\left(1-r\right)q_y\right\rbrace,
\end{align}
where a strain-induced vector potential can be read off as {$\bm {\mathcal A}^\xi= \xi \tfrac{4c}{p} (1+r)y\hat x$}, giving rise to a uniform strain-induced pseudo-magnetic field {$\bm{\mathcal B}^\xi=\xi\tfrac{4c}{p}(1+r)\hat{z}$}. The resulting strain-induced pseudo-Landau levels read
%as
\begin{align}
E_n=\pm t \sqrt{n|c|8(1-r^2)},\qquad n=0, 1, 2,\cdots,
\end{align}
which are dispersionless because the contribution from the inhomogeneous Fermi velocity is neglected. We are also able to solve the full eigenvalue problem of Eq.~(\ref{Heff_tri}) and obtain the dispersive pseudo-Landau levels (see Appendix~\ref{a3} for details)
\begin{align}\label{pLL_tri}
E_n^\xi=\pm t\sqrt{n|c|8\left[1-r^2+\xi pq_x(1-r)\right]},
\end{align}
whose validity is justified by its good match to the numerical bands obtained by diagonalizing the nearest-neighbor tight-binding model on the staggered zero-flux square lattice with triaxial strain [Figs.~\ref{fig7}(c) and~\ref{fig7}(d)].

\section{Connection with the honeycomb lattice}
\label{sec7}
\begin{figure}[t]
\centering
\includegraphics[width=8.8cm]{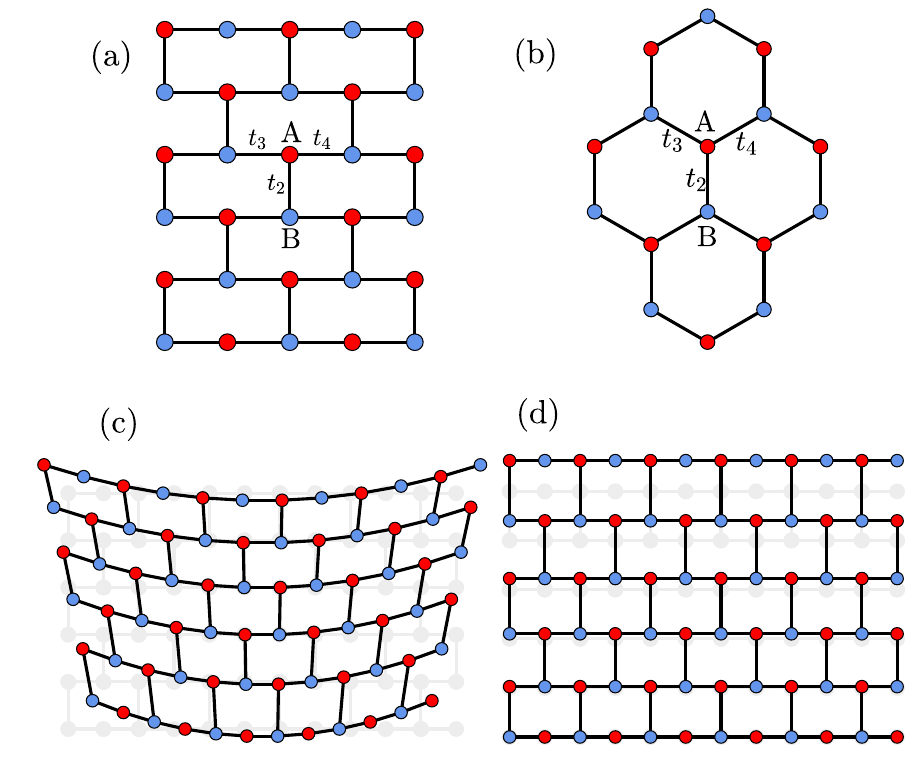}
\caption{Schematic plot of the staggered zero-flux square lattice with $t_1=0$ and the honeycomb lattice. The former is also known as the brick-wall lattice and is topologically equivalently to the latter. (a) Strain-free brick-wall lattice. (b) Strain-free honeycomb lattice. (c) Triaxially strained brick-wall lattice. (d) Uniaxially strained brick-wall lattice.
}\label{fig8}
\end{figure}

{In Sec.~\ref{sec6}, we have shown that the staggered zero-flux square lattice hosts a pair of Dirac cones when $0\leq t_1 <t$. In this section, we focus on the staggered zero-flux square with $t_1=0$ and study its connection with the honeycomb lattice. }
{We first investigate the spectrum of a nearest-neighbor tight-binding model on the staggered zero-flux square lattice. Breaking the $t_1$ bond (i.e., the flux-carrying bond in the staggered $\pi$-flux square lattice), the staggered zero-flux square lattice is reduced to the so-called ``brick-wall'' lattice [Fig.~\ref{fig8}(a)], which is topologically equivalent to a honeycomb lattice [Fig.~\ref{fig8}(b)] because they can be transformed into one another through continuous lattice geometry variation. The band structure of the brick-wall lattice reads}
\begin{equation}\label{E_bw}
  E_{\bm k}=\pm t\sqrt{3+2\cos(2k_x)+4\cos(k_x)\cos(k_y)},
\end{equation}
which is derived from the spectrum of Eq.~(\ref{H0k_sta}) upon setting $t_1=0$. There are two Dirac points at ${\bm K}^\xi=(\xi\frac{2\pi}{3},0)$, around which the low-energy effective Hamiltonian writes as $H^\xi_{\bm q}=t(-\xi\sqrt{3}\sigma_x q_x+\sigma_y q_y)$, implying an anisotropic Dirac cone. In contrast, the band structure of the corresponding honeycomb lattice~\cite{castroneto2009} is given by
\begin{equation}\label{E_hc}
  E^{\rm h}_{\bm k}=\pm t\sqrt{3+2\cos\left(\sqrt{3}k_x\right)+4\cos\left(\frac{\sqrt3}2k_x\right)\cos\left(\frac32k_y\right)}.
\end{equation}
Equation~(\ref{E_hc}) also exhibits two Dirac points $\bm K^{\text h,\xi}=(\xi\tfrac{4\pi}{3\sqrt 3},0)$, around which the low-energy effective Hamiltonian writes as $H^{\text h,\xi}_{\bm q}=\tfrac{3}{2}t(-\xi\sigma_x q_x+\sigma_y q_y)$, implying an isotropic Dirac cone. It is thus evident that the brick-wall lattice is linked to the honeycomb lattice under the following transformation
\begin{equation}\label{trans}
k_x\rightarrow\frac{\sqrt3}2k_x,\ k_y\rightarrow\frac32k_y,
\end{equation}
which corresponds to the lattice geometry variation from the brick-wall lattice to the honeycomb lattice.

We next consider the strain effect. Here, our focus on the non-uniform triaxial strain, and the case of non-uniform uniaxial strain is similar. The low-energy effective Hamiltonian of the brick-wall lattice under the triaxial strain can be obtained directly by setting $r=0$ in Eq.~(\ref{Heff_tri_simp}). It reads
\begin{align}\label{Hs_bw}
  H_{\bm q}^\xi&=t\left\{\sigma_x\left[-\xi {\sqrt3}q_x+4cy\right]+\sigma_yq_y\right\}.
\end{align}
For the honeycomb lattice, the effective Hamiltonian under the triaxial strain can be written as~\cite{sun2022}
\begin{align} \label{Hs_hc}
  H_{\bm q}^\xi&=\frac{3}2t\left\{\sigma_x[-\xi q_x+2cy]+\sigma_y[q_y+\xi2cx]\right\}.
\end{align}
Comparing Eq.~(\ref{Hs_bw}) to Eq.~(\ref{Hs_hc}), we find that the lattice geometry influences both the Fermi velocity and the strain-induced pseudo-vector potential. On the one hand, the Fermi velocities of the two lattices are connected by the transformation Eq.~(\ref{trans}). On the other hand, the pseudo-vector potentials of the two lattices are different in gauge. While the pseudo-vector potential of the honeycomb lattice, ${{\bm {\mathcal A}}^\xi=\xi(2cy,-2cx)}$, is symmetric, the pseudo-vector potential of the brick-wall lattice only contains a non-zero $x$ component ${\cal A}^{\xi}_x=4\xi cy/\sqrt{3}$. The disappearance of ${\cal A}^{\xi}_y$ is attributed to the square geometry, where $t_3$ and $t_4$ are always equal under strain [Eq.~(\ref{t_mod_tri})]. Although lattice geometry variation leads to discrepancy in the effective Hamiltonians, the resulting pseudo-Landau levels can still be expressed using a general formula
\begin{align}
E_n=\pm \hbar v_{F} \sqrt{8|c|n},\qquad n=0, 1, 2,\cdots,
\end{align}
where {$v_{F}=t/\hbar$ $(v_F=3t/2\hbar)$} for the brick-wall and honeycomb lattices, respectively.

Lastly, it is worth noting that the translational symmetry is completely broken by the triaxial strain in the honeycomb lattice. However, in the case of the brick-wall lattice, the translational symmetry along the $x$ direction is preserved, because the strain-induced pseudo-vector potential adopts a Landau gauge [Eq.~(\ref{Hs_bw})]. This is evident from the fact that the hopping parameters, which are modulated by the triaxial strain, depend only on the $y$ coordinate [Eq.~(\ref{t_mod_tri})].

\section{Conclusions}
\label{sec8}
We have investigated the strain effects on columnar and staggered $\pi$-flux square lattices. Our analysis using low-energy effective theory reveals that strain applied to these $\pi$-flux square lattices does not induce pseudo-magnetic fields, but rather leads to inhomogeneous Fermi velocities. We further explore alternative methods to generate pseudo-magnetic fields in these systems. For columnar $\pi$-flux square lattices, pseudo-magnetic fields can be created through non-uniform on-site potentials, while for staggered $\pi$-flux square lattices, pseudo-magnetic fields require anisotropic hoppings. We have validated the theoretical predictions of the pseudo-Landau levels through numerical simulations using corresponding nearest-neighbor tight-binding models. The difference between pseudo-Landau levels and ordinary Landau levels arising from magnetic fields are discussed.

Removing the flux from the staggered $\pi$-flux square lattice, we find both uniaxial and triaxial strain can induce pseudo-magnetic fields and dispersive pseudo-Landau levels. Additionally, the strain-free and strained staggered zero-flux square lattices should in principle be more experimentally feasible than their $\pi$-flux counterparts. Further breaking the $t_1$ bond (i.e., the flux-carrying bond in the staggered $\pi$-flux square lattice) of the staggered zero-flux square lattice, we find that the resulting brick-wall lattice is topologically equivalent to the honeycomb lattice. While the strain-free band structures of these two lattices can be made equivalent by stretching and shrinking the Brillouin zone, the pseudo-magnetic fields generated under the same triaxial strain are different in gauge. The triaxial-strain-deformed honeycomb lattice has a symmetric pseudo-vector potential, while the pseudo-vector potential in the triaxial-strain-deformed brick-wall lattice aligns along the $x$ direction.
These results expand the effect of strain to square geometries, further enhancing our comprehension of the strain-induced pseudo-magnetic field. It is very possible that our strained lattices could be experimentally implemented in artificial platforms, such as optical lattices~\cite{aidelsburger2013, miyake2013} or electrical circuits~\cite{lee2018}.

Recently, metamaterials have been instrumental in the study of pseudo-magnetic fields and associated physical phenomena in honeycomb lattices. Experimental progress has been made in observing pseudo-Landau levels on various artificial platforms\cite{PhysRevLett.127.136401,RN60,PhysRevB.88.115437,RN62}, such as photonic\cite{RN61,RN63,RN64}, phononic\cite{doi:10.1073/pnas.1615503114,RN59,PhysRevLett.118.194301},and topolectric\cite{PhysRevLett.124.046401,teo2023pseudomagnetic} systems. Currently, there is no ideal two-dimensional quantum material with a square lattice structure similar to graphene that can perfectly replicate the honeycomb lattice. However, we anticipate that our theoretical findings can be validated through the utilization of artificially strained square lattices engineered using the aforementioned metamaterials. As an example, the two essential ingredients required to engineer the pseudo-magnetic field, namely spatially non-uniform on-site potentials and anisotropic hoppings, can be effectively induced in topolectric circuits composed of simple elements such as capacitance and inductance\cite{PhysRevLett.122.247702,PhysRevResearch.3.023056}. The positive hoppings are achieved through the use of capacitors, while the negative hoppings are achieved by carefully selecting the appropriate inductors. The spatial variations in the hoppings required to generate the pseudo-magnetic field can be created by connecting additional suitable inductors and capacitors. The on-site chemical potentials can be manipulated by applying distinct grounding elements to each site. Furthermore, a recent proposal suggests that a tight-binding model with arbitrary hopping amplitudes and phases can be constructed by extending a node in an LC circuit\cite{PhysRevResearch.3.023056}. Given the rapid advancements in topolectric circuits, it is highly likely that our theoretical results pertaining to strained square lattices will be observed experimentally.

\appendix

\section{{Non-uniform potential induced pseudo-Landau levels in the columnar $\pi$-flux square lattice}}
\label{a1}
In Sec.~\ref{sec4} of the main text, we have mentioned that a non-uniform potential produces pseudo-Landau levels [Eq.~(\ref{LL_col})] in the columnar $\pi$-flux square lattice. We now explicitly derive Eq.~(\ref{LL_col}) by solving the eigenvalue problem of Eq.~(\ref{Hq_col}).

Writing the wave function as $\Psi=e^{iq_yy}(\phi_A,\ \phi_B)^T$, the eigenvalue problem of Eq.~(\ref{Hq_col}) explicitly reads
\begin{equation} \label{ep}
\begin{split}
\left(\xi q_y +{U_0}x-\epsilon \right)\phi_A =\frac{\partial }{\partial x}\phi_B,
\\
\left(\xi q_y +{U_0}x+\epsilon \right)\phi_B =\frac{\partial }{\partial x}\phi_A,
\end{split}
\end{equation}
where we define $\epsilon=\frac{E}{2t}$. Equation~(\ref{ep}) can be rewritten as
\begin{equation} \label{epf}
\begin{split}
\left(\xi q_y + {U_0}x-\frac{\partial }{\partial x}\right)f_1 =\epsilon f_2,
\\
\left(\xi q_y + {U_0}x+\frac{\partial }{\partial x}\right)f_2 =\epsilon f_1,
\end{split}
\end{equation}
where we define new variables $f_1=\phi_A+\phi_B$ and $f_2=\phi_A-\phi_B$. Since ${[\xi q_y + U_0x+\frac{\partial }{\partial x}, \xi q_y + U_0x-\frac{\partial }{\partial x}] = 2U_0}$, we can define the following bosonic ladder operators
\begin{equation}
\begin{split}
\beta=\frac{1}{\sqrt{2{U_0}}}\left(\xi q_y +{U_0}x+\frac{\partial }{\partial x}\right),
\\
\beta^\dagger=\frac{1}{\sqrt{2{U_0}}}\left(\xi q_y +{U_0}x-\frac{\partial }{\partial x}\right).
\end{split}
\end{equation}
Making use of Eq.~(\ref{epf}), we obtain
\begin{equation}
2{U_0}\beta^{\dagger } \beta f_2 =\epsilon^2 f_2.
\end{equation}
Note that $\beta^\dagger\beta$ is a bosonic number operator whose eigenvalues are $n=0,1,2,\cdots$. We thus have $\epsilon=\pm\sqrt{2 {U_0} n}$ and the resulting pseudo-Landau levels read
\begin{equation}
E_n=\pm2t\sqrt{2{U_0}n},\qquad n=0,1,2,\cdots,
\end{equation}
which is labeled as Eq.~(\ref{LL_col}) in the main text.

\section{{Anisotropic hoppings induced pseudo-Landau levels in the staggered $\pi$-flux square lattice}}
\label{a2}
In Sec.~\ref{sec5} of the main text, we have mentioned that  inhomogeneous anisotropic hoppings produce pseudo-Landau levels [Eq.~(\ref{LL_sta})] in the staggered $\pi$-flux square lattice. We now explicitly derive Eq.~(\ref{LL_sta}) by solving the eigenvalue problem of Eq.~(\ref{Hq_sta}).

For simplicity, we perform a variable substitution $x\rightarrow x+\frac2c$ to Eq.~(\ref{Hq_sta}). The resulting low-energy effective Hamiltonian becomes
\begin{equation}\label{Hq_sta_vsub}
\mathcal{H}^\xi_{\bm{q}}=2t\left[-\frac{c}{2}\xi x{\rm i}\partial_x\sigma_x+\left(q_y-\xi-\xi\frac{c}{2}x\right)\sigma_y\right].
\end{equation}
As ${\rm i}x\partial_x$ in Eq.~(\ref{Hq_sta_vsub}) is non-hermitian, we need to replace it with a hermitian operator, {${\rm i}x\partial_x\rightarrow {\rm i}(x\partial_x+\partial_xx)/2$}. Writing the wave function as $\Psi=e^{iq_yy}(\phi_A, \phi_B)^T$, the eigenvalue problem of the hermitianized effective Hamiltonian reads
\begin{equation}
\begin{split}
\left[ -i\frac{c}{2}\xi\left(x\partial_x +\frac{1}{2}\right)-i\left(q_y -\xi-\xi\frac{c}{2}x\right)\right] \phi_B =\epsilon \phi_A,
\\
\left[ -i\frac{c}{2}\xi\left(x\partial_x +\frac{1}{2}\right)+i\left(q_y -\xi-\xi\frac{c}{2}\;x\right)\right] \phi_A =\epsilon \phi_B,
\end{split}
\end{equation}
where we have defined $\epsilon=\frac{E}{2t}$. The elimination of $\phi_A$ gives rise to a second-order ordinary differential equation with respect to $\phi_B$ as
\begin{equation}\label{ode}
\begin{split}
\left[ x^2 -\frac{4\left(\xi q_y -1\right)-c}{c}x+\frac{\Delta }{c^2 } - \frac{1}{4}\right] \phi_B
\\
-\left(2x\phi_B^{\prime } +x^2 \phi_B^{\prime \prime } \right)=0,
\end{split}
\end{equation}
where we define {$\Delta =4 [(\xi q_y -1)^2 -\epsilon^2]$} for transparency.

We can then find the asymptotic solutions for $x\rightarrow 0$ and $x\rightarrow \pm\infty$, respectively. When approaching $x=0$, we can neglect the $x^2\phi_B$ and $x\phi_B$ terms in Eq.~(\ref{ode}), leading to the following asymptotic ordinary differential equation
\begin{equation}
\left( \frac{\Delta }{c^2 }-\frac{1}{4}\right) \phi_B -\left(2x\phi_B^{\prime } +x^2 \phi_B^{\prime \prime } \right)=0,
\end{equation}
which is a Cauchy-Euler equation with convergent solution $\phi_B\approx x^{-\frac12+\frac{\sqrt{\Delta}}{|c|}}$. As $x\rightarrow\pm\infty$, only $x^2\phi_B$ and $x^2 \phi_B^{\prime\prime }$ should be kept, leading to the following asymptotic ordinary differential equation
\begin{equation}\label{ode_inf}
\phi_B - \phi_B^{\prime \prime } =0.
\end{equation}
It is worth noting that the general solution to Eq.~(\ref{ode_inf}), $\phi_B=Ae^x+Be^{-x}$, cannot converge simultaneously as $x\rightarrow -\infty$ and $x\rightarrow+\infty$. To guarantee the $\pi$ flux, we require the engineered hopping $t_3$ to be positive. According to Eq.~(\ref{t_eng}), this requirement mandates that the solution must converge as {$x\rightarrow -\infty$ ($x\rightarrow +\infty$) if $c>0$ ($c<0$)}. Therefore, the convergent solution can be written as $\phi_B=e^{{\rm sgn}(c) x}$. With this \emph{a posteriori} solution, we can express the full solution to Eq.~(\ref{ode}) as $\phi_B=e^{{\rm sgn}(c) x}x^{-\frac12+\frac{\sqrt{\Delta}}{|c|}}u(x)$, which, upon substitution into Eq.~(\ref{ode}), yields
%%
%\begin{align}\label{eqB8}
%\nonumber
%  &-\frac{e^{{\rm sgn}(c) x}}{|c|} x^{\frac{1}{2}+\frac{\sqrt{\Delta }}{|c|}}\left\lbrace 2{\rm sgn}(c)\left(\sqrt{\Delta }+2\xi q_y +\frac{|c|-c}{2}-2\right) u\right. \\
%  &\left.+\left(|c|+2\sqrt{\Delta }+2cx\right)u^{\prime }+|c|xu^{\prime \prime } \right\rbrace =0
%\end{align}
%%
%By defining
%\begin{equation}
%  \gamma=1+\frac{2\sqrt\Delta}{|c|},\ \alpha=\frac1{|c|}(\sqrt\Delta+2\xi q_y+\frac{|c|-c}{2}-2)
%\end{equation}
%and $z=-{\rm sgn}(c)2x$, then Eq(\ref{eqB8}) is reduced to
%\begin{equation}
%  zu^{\prime \prime } \left(z\right)+\left(\gamma -z\right)u^{\prime } \left(z\right)-\alpha u\left(z\right)=0
%\end{equation}
%
\begin{equation}\label{cfl}
zu^{\prime \prime } \left(z\right)+\left(\gamma -z\right)u^{\prime } \left(z\right)-\alpha u\left(z\right)=0,
\end{equation}
where we define for transparency, {$\gamma=1+2\sqrt\Delta/|c|$, $\alpha=\frac1{|c|}(\sqrt\Delta+2\xi q_y+\frac{|c|-c}{2}-2)$, and $z=-2{\rm sgn}(c)x$.} Equation~(\ref{cfl}) is a confluent hypergeometric equation with a regular singularity at $z = 0$ and can be solved by the series expansion method. One solution reads
\begin{equation}
u(z)=1+\frac{\alpha}{\gamma}\frac{z}{1!}+\frac{\alpha(\alpha+1)}{\gamma(\gamma+1)}\frac {z^2}{2!}+\cdots,
\end{equation}
where $\gamma\neq0,-1,-2,\cdots$. To make $u(z)$ a polynomial (and therefore finite) function, $\alpha$ should be zero or negative integers ({i.e., $\alpha=-\nu$ with $\nu=0,1,2,\cdots$}). This constraint leads to the following expression for the eigenenergy
\begin{equation}
E_n^\xi=\pm2t\sqrt{n|c|(1-\xi q_y)},
\end{equation}
where $n=0,1,2,\cdots$ ($n=1,2,\cdots$) for $c>0$ ($c<0$). This equation is labeled as Eq.~(\ref{LL_sta}) in the main text.

%It  is  noteworthy  that  Eq.(B13)   is derived by solving the differential equation for $\phi_B$. By eliminating $\phi_B$ using Eqs. (B3) and (B4), we can obtain the second-order differential equation satisfied by $\phi_A$:
%\begin{equation}
%  \left\lbrace x^2 -\frac{4\left(\xi q_y -1\right)+c}{c}x+\frac{\Delta }{c^2 }-\frac{1}{4}\right\rbrace \phi_A -\left(2x\phi_A^{\prime } +x^2 \phi_A^{\prime \prime } \right)=0.
%\end{equation}
%Following  the  same  procedure  as  before,  we  can  solve  Eq(B14)  and  obtain  the  eigenvalues  of  $\phi_A$,  which  is
%\begin{equation}
%  E_n^\xi=\pm2t\sqrt{n|c|(1-\xi q_y)}
%\end{equation}
%for $c>0$($c<0$),we have $n=1,2,\cdots$($n=0,1,2,\cdots$).

\section{{Strain-induced pseudo-Landau levels in the staggered zero-flux square lattice}}
\label{a3}
In Sec.~\ref{sec6} of the main text, we have mentioned that both uniaxial and triaxial strain produce pseudo-Landau levels [Eqs.~(\ref{pLL_uni}) and~(\ref{pLL_tri})] in the staggered zero-flux square lattice. We now explicitly derive Eqs.~(\ref{pLL_uni}) and~(\ref{pLL_tri}) by solving the eigenvalue problems of Eqs.~(\ref{Heff_uni}) and~(\ref{Heff_tri}), respectively.

We start from the eigenvalue problem of Eq.~(\ref{Heff_uni}). For simplicity, we perform a variable substitution $y\rightarrow y+\frac{1-r-\xi pq_x}{2c}$, where $p=\sqrt{(3+r)(1-r)}$. The low-energy effective Hamiltonian [Eq.~(\ref{Heff_uni})] is changed to
\begin{equation}\label{Heff_uni_vsub}
{
H_{\bm q}^\xi = t\sigma_x\left[\frac{r-1}{2} \xi pq_x-(1+r)cy+\frac{r^2-1}{2}\right] + t\sigma_y {\rm i}2cy\partial_y.
}
\end{equation}
As the operator ${\rm i}y\partial_y$ in Eq.~(\ref{Heff_uni_vsub}) is non-hermitian, we again apply the replacement ${\rm i}y\partial_y\rightarrow {\rm i}(y\partial_y+\partial_yy)/2$ to restore the hermicity. Writing the wave function as $\Psi=e^{iq_xx}(\phi_A,\ \phi_B)^T$, the eigenvalue problem of the hermitianized effective Hamiltonian reads
%
%\begin{align}
%\nonumber
%  H_{\bm q}^\xi &= t\left\lbrace\sigma_x\left[-\xi \frac{1-r}{2}{\sqrt{(3+r)(1-r)}}q_x-(1+r)cy+\frac{r^2-1}{2}\right]\right.\\
%  &\left.+\sigma_y \left[{\rm i}2cy\partial_y+c\right]\right\rbrace.
%\end{align}
%
%
%\begin{align}
%&\left(-\xi aq_x-by+d+2cy\partial_y+c\right)\phi_B=\epsilon\phi_A,
%\\
%&\left(-\xi aq_x-by+d-2cy\partial_y-c\right)\phi_A=\epsilon\phi_B,
%\end{align}
%%
%where we have defined $a=\frac{1-r}{2}{\sqrt{(3+r)(1-r)}}, b=(1+r)c, d=(r^2-1)/2$ and $\epsilon=E/t$.
%
%
\begin{equation}
\begin{split}
{
\left[\frac{r-1}{2} \xi p q_x-(1+r)cy + \frac{r^2-1}{2} +2cy\partial_y+c\right]\phi_B=\epsilon\phi_A,
}
\\
{
\left[\frac{r-1}{2} \xi pq_x-(1+r)cy+ \frac{r^2-1}{2} -2cy\partial_y-c\right]\phi_A=\epsilon\phi_B,
}
\end{split}
\end{equation}
where we have defined $\epsilon=E/t$. By eliminating $\phi_A$, we obtain a second-order ordinary differential equation with respect to $\phi_B$ as
%
%\begin{align}\label{EqC5}
%\nonumber
%&\phi_B\left[\frac{\Delta}{4c^2}-\frac{1}{4}+\frac{b(c-d+\xi aq_x)}{2c^2}y+\frac{b^2}{4c^2}y^2\right]\\
%&-y\left(2 \phi^\prime_B+y \phi_B^{\prime\prime}\right)=0
%\end{align}
%
\begin{align}\label{ode_uni}
&{\left\{\frac{(1+r)^2}{4}y^2 + \frac{(1+r)[2c-r^2+1 + (1-r) \xi pq_x]}{4c}y \right.} \nonumber
\\
&{\left.+\frac{\Delta}{4c^2}-\frac{1}{4} \right\}\phi_B -\left(2 y\phi^\prime_B+y^2 \phi_B^{\prime\prime}\right)=0},
\end{align}
where we define {$\Delta=\frac{1}{4}[r^2-1-(1-r)\xi pq_x]^2-\epsilon^2$} for transparency.

Following the same strategy in Appendix~\ref{a2}, we first find the asymptotic solutions to Eq.~(\ref{ode_uni}). For $y\rightarrow0$, we neglect the $y^2\phi_B$ and $y\phi_B$ terms in Eq.~(\ref{ode_uni}), resulting in the following asymptotic ordinary differential equation
\begin{align}
&\left(\frac{\Delta}{4c^2}-\frac{1}{4}\right)\phi_B - \left(2 y\phi^\prime_B+y^2 \phi_B^{\prime\prime}\right)=0,
\end{align}
whose convergent solution is $\phi_B\approx y^{-\frac12+\frac{\sqrt{\Delta}}{2|c|}}$. As {$y\rightarrow\pm\infty$}, Eq.~(\ref{ode_uni}) is simplified as
\begin{align}
{
\frac{(1+r)^2}{4} \phi_B-  \phi_B^{\prime\prime}=0,
}
\end{align}
whose convergent solution is $\phi_B=e^{{\rm sgn}(c)\frac{(1+r)}{2}y}$. Consequently, the full solution to Eq.~(\ref{ode_uni}) reads $\phi_B=y^{-\frac12+\frac{\sqrt{\Delta}}{2|c|}}e^{{\rm sgn}(c)\frac{(1+r)}{2}y}u(y)$, which, upon plugging back into Eq.~(\ref{ode_uni}), leads to the following confluent hypergeometric equation
%\begin{align}
%\nonumber
%&c y u''(y)+u'(y) \left(|c| (r+1) y+{\rm sgn}(c) \sqrt{\Delta}+c\right)\\ \nonumber
%&+u(y)(r+1) \left(\frac{\sqrt{\Delta}}{2}-\xi q_x \frac{1-r}{4}\sqrt{(1-r)(r+3)} \right.\\
%&\left.+\frac{r^2-1}{4}+\frac{|c|-c}{2}\right)=0
%\end{align}
%
\begin{align}
zu''(z)+(\gamma-z)u'(z)-\alpha u(z)=0,
\end{align}
where $\gamma=1-\frac{2\sqrt\Delta}{|c|}$, $\alpha=\frac{1}{|c|}(\sqrt{\Delta}-\xi p q_x  +\frac{r^2-1}{2}+\frac{|c|-c}{2})$, and $z=-{\rm sgn}(c)(r+1)y$. As discussed in Appendix~\ref{a2}, we require $\alpha=0,-1,-2,\cdots$. This constraint requires the following eigenenergy
\begin{align}
{
E_n^\xi=\pm t\sqrt{2(1-r)}\sqrt{n|c|\left(\xi p q_x+1+r\right)},
}
\end{align}
which is labeled as Eq.~(\ref{pLL_uni}) in the main text.

We now turn to study the eigenvalue problem of Eq.~(\ref{Heff_tri}). For simplicity, we perform the variable substitution $y\rightarrow y-\frac{1-r-\xi pq_x}{2c(3+r)}$, where $p=\sqrt{(3+r)(1-r)}$. The low-energy effective Hamiltonian [Eq.~(\ref{Heff_tri})] is then mapped to
{
\begin{align} \label{Heff_tri_vsub}
H_{\bm q}^\xi=&t\sigma_x\bigg[\frac{(\xi pq_x+r-1)^2-p^2}{3+r} +2c(2+2r+\xi pq_x)y\bigg] \nonumber
\\
&+t\sigma_y2c(3+r)(-{\rm i}y\partial_y).
\end{align}
}
To restore the hermicity of Eq.~(\ref{Heff_tri_vsub}), we again apply the replacement ${\rm i}y\partial_y\rightarrow {\rm i}(y\partial_y+\partial_yy)/2$. Writing the wave function as $\Psi=e^{iq_xx}(\phi_A,\ \phi_B)^T$, the eigenvalue problem of the hermitianized effective Hamiltonian reads
{
\begin{equation}
\begin{split}
\bigg[\frac{(\xi pq_x+r-1)^2-p^2}{3+r} +2c(2+2r+\xi pq_x)y
\\
-2c(3+r)y\partial_y-c(3+r)\bigg]\phi_B=\epsilon\phi_A,
\\
\bigg[\frac{(\xi pq_x+r-1)^2-p^2}{3+r} +2c(2+2r+\xi pq_x)y
\\
+2c(3+r)y\partial_y+c(3+r)\bigg]\phi_A=\epsilon\phi_B,
\end{split}
\end{equation}
}
where we have defined $\epsilon=E/t$. By eliminating $\phi_A$, we arrive at a second-order ordinary differential equation with respect to $\phi_B$ as
\begin{equation} \label{ode_tri}
\begin{split}
&\left\{\frac{(2+2r+\xi pq_x)^2}{(3+r)^2}y^2+\frac{(2+2r+\xi pq_x)}{(3+r)c}\right.
\\
&\times \left[\frac{(\xi pq_x+r)^2+r^2}{(3+r)^2}+c-\frac{2(\xi pq_x+1)}{(3+r)^2}\right]y
\\
&\left.+\frac{\Delta}{4c^2(3+r)^4}-\frac{1}{4}\right\}\phi_B-(y^2\phi_B^{\prime\prime}+2y\phi_B^\prime)=0,
\end{split}
\end{equation}
where {$\Delta={\left[(\xi pq_x+r-1)^2-p^2\right]^2-(3+r)^2\epsilon^2}$} is defined for transparency.

Following the procedure solving Eqs.~(\ref{ode}) and~(\ref{ode_uni}), we study the asymptotic solutions of Eq.~(\ref{ode_tri}). For $y\rightarrow 0$, the terms associated with $y^2\phi_B$ and $y\phi_B$ in Eq.~(\ref{ode_tri}) can be safely neglected, resulting in the following asymptotic ordinary differential equation
\begin{align}
&\left[\frac{\Delta}{4c^2(3+r)^4}-\frac{1}{4}\right]\phi_B-\left(2y \phi^\prime_B+y^2 \phi_B^{\prime\prime}\right)=0,
\end{align}
whose convergent solution reads $\phi_B\approx y^{-\frac12+\frac{\sqrt{\Delta}}{2|c|(3+r)^2}}$. As ${y\rightarrow\pm\infty}$, Eq.~(\ref{ode_tri}) is reduced to
\begin{align}
{
\frac{(2+2r+\xi pq_x)^2}{(3+r)^2}\phi_B - \phi_B^{\prime\prime}=0,
}
\end{align}
whose convergent solution is {$\phi_B=e^{\text{sgn}(c)\frac{(2+2r+\xi pq_x)}{3+r}y}$}. The full solution of Eq.~(\ref{ode_tri}) can thus be written as {$\phi_B=y^{-\frac12+\frac{\sqrt{\Delta}}{2|c|(3+r)^2}}e^{\text{sgn}(c)\frac{(2+2r+\xi pq_x)}{3+r}y}u(y)$}, which, upon plugging back into Eq.~(\ref{ode_tri}), leads to the following confluent hypergeometric equation
\begin{equation}
zu''(z)+(\gamma-z)u'(z)-\alpha u(z)=0,
\end{equation}
{where $\gamma=[\sqrt\Delta+c^2(3+r)^2]/[2c(3+r)(2+2r+\xi pq_x)]$, $\alpha=[\sqrt{\Delta}-2(\xi pq_x+1)+2c(3+r)^2+(\xi pq_x+r)^2+r^2]/[2c(3+r)^2]$, and $z=-{\rm sgn}(c)\frac{2(2+2r+\xi pq_x)}{3+r}y$}. As discussed above, $\alpha$ should adopt zero or negative integers. This constraint requires the following eigenenergy
\begin{align}
E_n^\xi=\pm t\sqrt{n|c|8\left[1-r^2+\xi pq_x(1-r)\right]}
\end{align}
which is labeled as Eq.~(\ref{pLL_tri}) in the main text.

\begin{acknowledgments}
The authors are indebted to E. Lantagne-Hurtubise, X.-X. Zhang, M. Franz, Z. Shi, and H.-Z. Lu for insightful discussions. J.S and H.G. acknowledges support from the NSFC grant Nos.~11774019 and 12074022. T.L. acknowledges Guangdong Basic and Applied Basic Research Foundation under grant No. 2022A1515111034. S.F. is supported by the National Key Research and Development Program of China under grant No. 2021YFA1401803, and NSFC under Grant Nos. 11974051 and 12274036.
\end{acknowledgments}

\bibliographystyle{apsrev4-1-etal-title_10authors}
\bibliography{pmf20231006}

\end{document}